\documentclass[12pt,preprint]{aastex}
\slugcomment{To appear in the Astrophysical Journal}

\lefthead{Skinner et al.}
\righthead{CQ Cep}

\begin{document}

\title{A Chandra Observation of the Eclipsing Wolf-Rayet \\
  Binary CQ Cep}

\author{Stephen L. Skinner\footnote{CASA, Univ. of Colorado,
Boulder, CO, USA 80309-0389; stephen.skinner@colorado.edu},
Svetozar A. Zhekov\footnote{Space Research and Technology Institute, Akad. G. Bonchev Str.,
Sofia, 1113, Bulgaria; szhekov@space.bas.bg},
Manuel  G\"{u}del\footnote{Dept. of Astrophysics, Univ. of Vienna,
T\"{u}rkenschanzstr. 17,  A-1180 Vienna, Austria; manuel.guedel@univie.ac.at},
and
Werner Schmutz\footnote{Physikalisch-Meteorologisches Observatorium Davos and
World Radiation Center (PMOD/WRC), Dorfstrasse 33, CH-7260 Davos Dorf, Switzerland;
werner.schmutz@pmodwrc.ch}
}



%
\newcommand{\ltsimeq}{\raisebox{-0.6ex}{$\,\stackrel{\raisebox{-.2ex}%
{$\textstyle<$}}{\sim}\,$}}
%
\newcommand{\gtsimeq}{\raisebox{-0.6ex}{$\,\stackrel{\raisebox{-.2ex}%
{$\textstyle>$}}{\sim}\,$}}

\begin{abstract}
The  short-period (1.64 d) near-contact eclipsing WN6$+$O9 binary  system CQ Cep provides
an ideal laboratory for testing the predictions of X-ray colliding wind shock theory
at close separation where the winds may not have  reached terminal speeds before colliding.
We present results of a  {\em Chandra} X-ray observation of CQ Cep spanning $\sim$1 day 
during which a simultaneous {\em Chandra} optical light curve was acquired. Our primary 
objective was to compare the observed X-ray properties with colliding wind shock theory,  
which predicts that the hottest shock plasma (T $\gtsimeq$ 20 MK) will form  on or near 
the line-of-centers between the stars. The X-ray spectrum is strikingly similar to 
apparently single WN6 stars such as WR 134 and spectral lines reveal plasma over a broad
range of temperatures T $\sim$ 4 - 40 MK.  A deep optical eclipse was seen as the O star 
passed  in front of the Wolf-Rayet star and we determine an orbital period 
P$_{orb}$ = 1.6412400 d. Somewhat surprisingly, no significant X-ray  variability was detected.
This implies that the hottest X-ray plasma is not confined to the region between the stars, at 
odds with  the colliding wind picture and suggesting that other X-ray production mechanisms
may be at work. Hydrodynamic simulations  that account
for such effects  as radiative cooling and orbital motion will be needed to determine if the 
new {\em Chandra} results can be reconciled with the colliding wind picture.
\end{abstract}


\section{Introduction}
Massive WR$+$O binary systems are  luminous X-ray sources
with typical luminosities L$_{x}$ $\sim$ 10$^{32.5}$ - 10$^{34}$ ergs s$^{-1}$.
The origin of their X-ray emission is complex and multiple processes may 
contribute. X-rays can arise in the powerful winds of either star  and
an additional component can originate in a colliding wind shock that forms
between the stars. 

X-ray emission from the O-star in WR$+$O binaries is expected since
single O stars are X-ray sources and their  X-ray luminosity is
correlated with bolometric luminosity  L$_{x}$ $\sim$ 10$^{-7}$L$_{bol}$ 
(Pallavicini et al. 1981; Bergh\"{o}fer et al. 1996; 1997). Soft X-ray
emission (kT $\ltsimeq$ 1 keV) in O stars is usually attributed to
radiative shocks that form in their winds as a result of line-driven
instabilities (Lucy \& White 1980; Owocki et al. 1988; Feldmeier et al. 1997).
Some observational support for this process has been found in
objects like the O4 star $\zeta$ Puppis (Cassinelli et al. 2001; Kahn et al. 2001).
But if the star has a strong magnetic field then harder X-rays  (kT $\gtsimeq$ 2 keV) 
may be produced in shocks near the magnetic equator via magnetic wind confinement 
(Babel \& Montmerle 1997;  ud-Doula \& Owocki 2002). 

In the case of WN$+$O binaries, the nitrogen-type WN star may also be
an X-ray source since several apparently single WN stars have been detected
in X-rays with typical luminosities  
L$_{x}$ $\sim$ 10$^{31.5}$ - 10$^{33}$  ergs s$^{-1}$ (Skinner et al. 2010; 2012).
Their X-ray spectra usually show both a soft and hard X-ray component.
The soft emission could arise in radiative wind shocks, similar to that 
theorized for single O stars (Gayley \& Owocki 1995). However, the origin of the 
hard X-ray component in putatively single WN stars is so far unexplained. Single 
carbon-type (WC) Wolf-Rayet stars have not yet been detected in X-rays with 
upper limits  as stringent as L$_{x}$ $\leq$ 10$^{29.8}$ ergs s$^{-1}$
(Skinner et al. 2006). 

Even though single WC stars have so far eluded X-ray detection, WC$+$O binaries
have been detected as strong X-ray sources as well as WN$+$O binaries. 
The most extensively studied WR$+$O binaries are the brightest in X-rays
and are believed to produce a substantial fraction of their X-ray emission
from colliding wind shocks, 
namely $\gamma^2$ Vel (WC8 $+$ O7.5; P$_{orb}$ = 78.5 d;
Willis, Schild, \& Stevens 1995; Skinner et al. 2001; Schild et al. 2004),
WR 140 (WC7 $+$ O4-5V; P$_{orb}$ = 2900 d; Zhekov \& Skinner 2000; Pollock et al. 2005),
and WR 147 (WN8 $+$ OB; Skinner et al. 2007; Zhekov \& Park 2010).
An X-ray study of short-period WR $+$ O  binaries (P$_{orb}$ $<$ 23 d) based on
archival data was presented by Zhekov (2012) who argued for the presence of a colliding 
wind shock component.

The X-ray emission of WR$+$O binaries is almost always unresolved with existing
telescopes and  typically includes both a soft and hard component.
At  least some of the hard emission (kT $\gtsimeq$ 2 keV)
is  thought to originate  in colliding wind shocks (Cherepaschuk 1976; 
Prilutski \& Usov 1976; Usov 1992). But for WN$+$O systems this may not be the 
whole story  since apparently single WN stars without known companions always
show a hard component in their spectra. Thus, the WN star itself may contribute
to the X-ray emission of WN$+$O binaries.

Disentangling the stellar and colliding wind  X-ray contributions from WR$+$O 
binaries based on the spectrum of a spatially unresolved source is challenging
because  the various contributions are superimposed in the composite spectrum.
High-resolution X-ray grating spectra can provide constraints on  where individual
emission lines are formed in the system but such spectra are only available
for a few of the brightest WR$+$O binaries.

Another approach to identifying separate X-ray contributions in WR $+$ O binaries
is to observe eclipsing systems. When the orbit is highly inclined (i.e. viewed
nearly edge-on), orbital modulation of the X-ray emission can occur. For example,
the hottest colliding wind shock plasma is expected to have temperatures of 
several keV for typical terminal wind speeds of  O-type and WR stars 
($V_{\infty}$ $\sim$ 1000 - 2000 km s$^{-1}$) and is predicted by numerical
hydrodynamic models to lie on the line-of-centers between the stars 
(Luo et al. 1990; Stevens et al. 1992). In that case, the hottest colliding
wind plasma will be visible near quadrature but will be partially occulted 
in high-inclination systems during 
primary and secondary eclipse. Thus, the observed emission measure of the hot
plasma component should vary with orbital phase.
Additionally, large variations in X-ray absorption
are anticipated with higher absorption occurring at orbital phases when the
WR star is in front of the O star and the system is viewed through the
dense metal-rich WR wind. Such orbital variations in absorption have indeed
been observed in well-studied binaries such as $\gamma^2$ Vel  and WR 140.

We present here the results of a {\em Chandra} observation of the short-period eclipsing
Wolf-Rayet binary CQ Cep covering more than half an orbit.
Our primary objectives were to search for X-ray modulation that should 
occur during eclipses if hot localized colliding wind shock plasma is
located on the line-of-centers and make comparisons between the observed 
X-ray properties and analytic predictions from colliding wind theory.
Somewhat surprisingly, no significant X-ray modulation was detected.

\section{CQ Cep}

The properties of our target   CQ Cep ( = WR 155 = HD 214419) are summarized in Table 1.   
It is a close binary system consisting of a WN6 nitrogen-type WR star and a O9  companion.
The luminosity class of the O9 companion remains rather uncertain because spectral features
clearly associated with the O star have been difficult to identify in optical spectra.
The O star was classified as  O9 II-Ib by Marchenko et al. (1995) but a O9 III-V
classification was suggested by Demircan et al.  (1997; hereafter De97). 
Harries \& Hilditch (1997) concluded that the inferred O star radius is consistent with
a  main sequence object.
The orbital period determined by De97 is P$_{orb}$ = 1.6412299 d (= 141.8 ks)
but slightly different values have been obtained in other studies (Sec. 5.5).  
The orbit is nearly
circular (Marchenko et al. 1995) and is viewed almost edge on with high inclination
$i$ $\geq$ 70$^{\circ}$ (De97; Harries \& Hilditch 1997; Villar-Sbaffi et al. 2005). 
The masses and radii of the 
two components are nearly equal and their separation is
$D$ $\approx$ 20.4 R$_{\odot}$, placing the two stars nearly in contact (De97).   
Because of the close separation, 
CQ Cep presents an opportunity to search for colliding wind shock emission in an
eclipsing system where the winds may not have reached terminal speeds before colliding.

CQ Cep was detected in X-rays by the {\em Einstein Observatory} (Pollock 1987)
and in a  1998 {\em ASCA} observation (sequence numbers 26024000,
26025000, 26026000; 45 ks total duration). Our fits of the archived {\em ASCA} spectra
show that both a cool
(kT$_{cool}$ $<$ 1 keV) and hot (kT$_{hot}$ $>$ 2 keV) plasma
component are present.  Although the {\em ASCA} data provide a useful first-look spectrum
and X-ray flux estimates, the signal-to-noise ratio is low and the X-ray properties
are not tightly constrained. In order to more accurately characterize the X-ray spectrum and 
search for orbital variability we obtained a more sensitive {\em Chandra} X-ray observation
along with a simultaneous  {\em Chandra} optical light curve.

\section{Chandra Observation}

The {\em Chandra} observation (ObsId 14538) began on
19 March 2013 at 12:56:12 TT (= 12:55:05 UTC)  and ended on 
20 March at 12:43:09 TT (= 12:42:02 UTC) with an exposure live time of 79,391 s.  
The start time corresponds to MJD = 56370.538252 (UTC) 
or HJD = 2456371.0353215 (UTC).  Based on the 
time of primary optical minimum given by  Demircan et al. (1997), the start time corresponds
to orbital phase $\phi$ = 0.92 and
the stop time to phase $\phi$ = 0.52, where $\phi$ = 0.0
corresponds to O star in front. However, the predicted orbital phases
corresponding to the {\em Chandra} start and stop times
depend on the  historical reference data used for optical minimum light,
as discussed further in Section 5.5. 
Exposures were obtained using the ACIS-S (Advanced CCD
Imaging Spectrometer) array in FAINT  timed-event mode.
CQ Cep was placed at the nominal aimpoint on the  ACIS-S3 CCD which
was configured  in 1/4 subarray mode. The use of subarray mode
restricts the field-of-view to a rectangular region of
256 $\times$ 1024 native pixels, or $\approx$126$''$ $\times$ 504$''$
at the native pixel size of 0.$''$492. Subarray mode was selected
in order to use a short 0.84 s frame time that would mitigate 
any photon pileup (none occurred). For an on-axis point source,
the ACIS-S 90\%  encircled energy radius at 1.4 keV  is
R$_{90}$ $\approx$ 0.$''$9. Optical monitoring of CQ Cep was
carried out during the observation using the Aspect Camera 
Assembly (ACA) and a photometric light curve was generated using the 
$monitor\_photom$ script.
Further information on {\em Chandra} and its instrumentation can
be found in the {\em Chandra} Proposer's
Observatory Guide (POG)\footnote {See http://asc.harvard.edu/proposer/POG}.

The pipeline-processed data  files provided by the {\em Chandra} X-ray
Center (CXC) were  analyzed using standard science
threads with CIAO version 4.6.1\footnote{Further information on
{\em Chandra} Interactive
Analysis of Observations (CIAO) software can be found at
http://asc.harvard.edu/ciao.}.
The CIAO processing  used recent calibration
data from CALDB 4.6.1.1. 
Source events, spectra, and light curves were extracted from a circular region of
radius 1.$''$8 centered on the X-ray peak. 
Background was extracted from nearby source-free  regions.
Background is negligible,
amounting to less than 1 count (0.2 - 8 keV) within the extraction
circle during the full exposure.
CIAO {\em specextract} was used to extract
spectra along with source-specific
response matrix files (RMFs) and auxiliary response files (ARFs).
Spectral fitting, timing analysis, and image analysis were undertaken with the HEASOFT
v. 6.1.6\footnote{http://heasarc.gsfc.nasa.gov/docs/xanadu/xanadu.html.}
software package  including XSPEC vers. 12.8.2.
Additional tests  for source variabilility
were carried out on  energy-filtered source event lists using
the Bayesian-method CIAO tool {\em glvary}
(Gregory \& Loredo 1992, 1996).

\section{Results}

CQ Cep was clearly detected. 
Its X-ray properties are summarized in Table 2.

\subsection{Image Analysis}

The broad-band X-ray image is shown in Figure 1.
The X-ray centroid is in excellent agreement with
the {\em Hubble Space Telescope} {\em Guide Star Catalog}
(GSC vers. 2.3.2) position of CQ Cep. In order to 
check for possible extended source structure,
we generated  a simulated   {\em Chandra}
point-spread-function (PSF) image using the 
{\em Chart} and {\em MARX}\footnote{http://space.mit.edu/ASC/MARX/}
software tools, following
recommended {\em CIAO} science  thread procedures.
The simulated PSF takes into account the position
of the source on the ACIS-S detector and the 
source spectrum. Using this PSF,  the CIAO $srcextent$
tool determines that the X-ray source is {\em not}
extended at 90\% confidence. 
Thus, CQ Cep is detected as a point source 
at {\em Chandra}'s spatial resolution.

\subsection{Timing Analysis}

Figure 2 shows the broad-band (0.2 - 8 keV) and hard-band (2 - 8 keV) 
X-ray light curves of CQ Cep along with the ACA optical light curve. 
No large-amplitude X-ray variability is seen but fluctuations
at the $\pm$2$\sigma$ level are present in the broad-band light curve.
The CIAO $glvary$ tool gives a probablility of constant count rate
P$_{const}$ = 0.77 (0.2 - 8 keV) and  $\chi^2$ analysis of the 
light curve binned at 2000 s intervals gives P$_{const}$ = 0.43 (0.2 - 8 keV).
In the soft band $glvary$ gives P$_{const}$ = 0.87 (0.2 - 2 keV)
and in the hard band  P$_{const}$ = 0.89 (2 - 8 keV).
Based on the above tests, variability is unlikely but not strictly ruled out.

As a further check for variability, we computed the median photon
energies (E$_{50}$) and hardness ratios (H.R.) for the first and second halves of
the observation. We define the hardness ratio as 
H.R. = (H $-$ S)/(H $+$ S) where H and S are the counts measured in
the hard (2 - 8 keV) and soft (0.2 - 2 keV) bands. For the first half we obtained
E$_{50}$ = 1.76 keV and H.R. = $-$0.280. For the second half,
E$_{50}$ = 1.77 keV and H.R. = $-$0.276. The values for both halves
are nearly identical.

The X-ray count rates based on light curves binned at 2000 s intervals
are given in Table 3 for three different energy bands. A slight 
increase in count rate in the second half of the observation is
seen in all three bands. The increase amounts to $\approx$10\% in
the broad (0.2 - 8 keV) and soft (0.2 - 2 keV) bands and 
$\approx$7\% in the hard (2 - 8 keV) band. Since the small increase is
present in all bands it could well be  real but the significance of
the increase is low, amounting  to at most 0.5$\sigma$ - 0.6$\sigma$.

In contrast, the optical light curve is clearly variable. A large drop in
brightness of 
$\approx$0.5 ACA mag\footnote{The zero instrument magnitude is defined as the 
ACA response to a zero magnitude G0V star. For the conversion  from V and B
magnitude to the ACA system see: http://cxc.harvard.edu/proposer/POG/html/chap5.html.} 
occurs during the first 15 ks with minimum brightness at
t = 14.46 $\pm$ 0.02 ks  after the start of the observation. This minimum corresponds to 
orbital phase $\phi$ = 0.023 using the time-of-minimum light equation of
De97:~HJD = 2450267.43158 $\pm $0.00077 $+$ (1.6412299 $\pm$ 0.0000002 d)$\cdot$E,
where the O star is in front of the WR star at $\phi$ = 0.0.
Thus, the minimum in the {\em Chandra} optical light curve occurred
$\approx$3.26 ks {\em after} the time predicted by the De97 elements (Fig. 2-top).
This difference cannot be explained by the small formal uncertainties in
the elements of De97. The uncertainty in their  HJD reference date 
could account for 66 s of the difference and their  orbital period uncertainty could
account for an additional 64 s when propagated forward to the epoch of
the {\em Chandra} observation (3718.92 orbits). This raises the question of whether 
the true uncertainties in the orbital  period of De97 may actually be larger than their quoted values or
whether the period might be variable. Both possibilities are  discussed further in Section 5.5.

\vspace*{0.6cm}

\subsection{Spectral  Analysis}

The {\em Chandra} ACIS-S spectrum of CQ Cep is shown in Figure 3.
Several emission lines and line blends are present including
high-temperature lines such as S XV at 2.46 keV (maximum line
power temperature log T$_{max}$ = 7.2 K),
Ar XVII at 3.13 keV (log T$_{max}$ = 7.3 K), Ca XIX at 3.90 keV
(log T$_{max}$ = 7.5 K), and a faint feature at 6.67 keV that
is likely weak Fe K$\alpha$  dominated by  Fe XXV (log T$_{max}$ = 7.6 K).
Emission features at lower energies that trace cooler plasma include
possible detections of the Ne IX He-like triplet at 0.91 keV (log T$_{max}$ = 6.6 K)
and Ne X at 1.02 keV (log T$_{max}$ = 6.7 K).
Thus, the X-ray plasma spans a broad range of temperatures from 
$\sim$4 MK up to $\sim$40 MK. Since the spectrum  becomes 
heavily absorbed at E $<$ 0.7 keV, any cooler plasma at 
T $<$ 4 MK that may be present would have escaped detection.  

The bottom panel of Figure 3 shows a comparison spectrum of the WN6 star
WR 134 based on a previous {\em Chandra} ACIS-S observation
(Skinner et al. 2010). It is noteworthy that its spectrum
is quite similar to that of CQ Cep despite the fact that 
WR 134 is not known to be a binary. This suggests that the 
WN star in CQ Cep dominates the observed X-ray emission. 

We have attempted to fit the CQ Cep spectrum with a variety of different
XSPEC models. We have focused on optically-thin plasma models ($apec$ and its
variable abundance version $vapec$) since the X-ray spectra of WN stars
and close WR$+$O binaries are generally well-fitted by such models 
(Skinner et al. 2010, 2012; Zhekov 2012). But we have also considered
a constant-temperature plane-parallel shock model ($phshock$ and its variable abundance version
$vpshock$)\footnote{Further information on XSPEC models can be found at:
http://heasarc.gsfc.nasa.gov/xanadu/xspec/manual/Models.html.}.
All models included a solar-abundance  absorption component
to account for interstellar medium (ISM) absorption and its value was held
fixed at N$_{H,ISM}$ = 2.2 $\times$ 10$^{21}$A$_{\rm V}$ (Gorenstein 1975)
= 4.4 $\times$ 10$^{21}$ cm$^{-2}$ corresponding to A$_{\rm V}$ = 2 mag.
We also allowed for additional absorption that is usually present
in WR star X-ray spectra as a result of their winds. Wind abundances
for the O star are expected to be near-solar in composition whereas
WN wind abundances are H-depleted and N-enriched  
(van der Hucht, Cassinelli, \& Williams 1986). Our fits
confirmed that excess absorption is present in CQ Cep and 
models with two plasma temperature components (2T) generally
provided better fits as determined  by the $\chi^2$ statistic
than models with only one temperature component (1T).

Table 4 summarizes three representative spectral fits using a 
2T optically thin plasma model and one fit using a 2T plane-parallel
shock model. Each model includes a fixed
ISM absorption component as well as  separate independent wind absorption 
components for the cool and hot plasma. In model A all abundances
were  assumed to be solar while in model B   the source
and wind-absorption abundances were held fixed at the generic
values for WN stars given by van der Hucht et al. (1986).
In model C the source and wind-absorption abundances of the cool
component were fixed at solar values (as expected for the O star)
and those for the hot component were fixed at generic WN values.
Model D uses the same abundance assumptions as model C but 
the 2T optically thin plasma model  is replaced by
a 2T plane-parallel shock model.

Even though the abundance assumptions for the  models in
Table 4 are different, the best-fit plasma temperatures (kT),
absorbed fluxes,  and unabsorbed X-ray luminosities (L$_{x}$) are 
all quite similar. Thus, the model results are not very sensitive
to abundance assumptions.
All models require a cool plasma component at 
kT$_{1}$ $\approx$ 0.6 keV and a hotter component at 
kT$_{2}$ $\approx$ 2 - 3 keV. These temperatures are comparable
to those found for other close WN$+$O binaries (Zhekov (2012)
as well as single WN stars (Skinner et al. 2010; 2012).
The models in Table 4 give log L$_{x}$ = 33.25 - 33.48
ergs s$^{-1}$ (0.3 - 8 keV) at d = 3.5 kpc. These values place CQ Cep at the
high end of the X-ray luminosity range observed for WR stars (Pollock 1987).
The above  L$_{x}$ range overlaps that of other close 
WN$+$O binaries (Fig. 2 of Zhekov 2012) but is also similar
to that of a few of the brightest putatively single WN6 stars 
such as WR 20b  and WR 24 (Skinner et al. 2010).

Of the models in Table 4, the 2T $vpshock$ model (model D) provides a
slightly better fit than the other models as gauged by $\chi^2$ statistics.
It also yields a somewhat higher
value of kT${_2}$ and L$_{x}$ than the 2T $vapec$ models. The slightly
better fit obtained with the $vpshock$ model is due to its ability to fully
reproduce the fluxes of the stronger emission line complexes, which are generally
underestimated by the $vapec$ models. This may be a clue that
non-equilibrium ionization (NEI) effects taken into account by the $vpshock$
model (Borkowski, Sarazin, \& Blondin 1994) are important but higher resolution 
grating spectra capable of resolving individual line complexes (e.g. He-like triplets) 
would be needed to confirm this. 

Finally, we note that the absorbed flux measured
from the best-fit $vpshock$ model applied to data from the first half
of the exposure is 4\% less than that obtained by fitting data from the second
half. Specifically, the absorbed fluxes measured from the first and second halves
in the 0.3 - 8 keV range and their 90\% confidence ranges are (in units of
10$^{-13}$ ergs cm$^{-2}$ s$^{-1}$):~
F$_{x,first}$ = 2.03 [0.80 - 2.25] and F$_{x,second}$ = 2.11 [0.93 - 2.32]. 
The small increase in measured flux provides additional support for the conclusion based
on count rates (Table 3) that the source was slightly brighter during the 
second half of the observation, but the 90\% confidence ranges of the 
measured fluxes are large enough to be consistent with no flux variability at all.

\section{Discussion}

\subsection{Colliding Wind Predictions}

\subsubsection{Wind Ram Balance and Stagnation Point}

In order to compare the observational results with  predictions
of colliding wind  theory we must first determine the location of
the stagnation point where the shock contact discontinuity surface
intersects the line-of-centers between the stars. As a first
approximation we assume radial outflow, neglect orbital motion (Sec. 5.2),
and consider one-dimensional wind momentum balance.
If the distance from the center of the O star to the stagnation point
is $r_{1}$ and  the distance from the center of the WR star to 
stagnation point is $r_{2}$  = $D$ - $r_{1}$, where $D$ is the binary
separation, then the necessary condition for 
wind ram balance is  (eq. [1] of Stevens et al. 1992)~

\begin{equation}
\frac{\dot{M}_{O9} V_{O9}(r_{1})}{\dot{M}_{WR} V_{WR}(r_{2})}  = \frac{r_{1}^2}{r_{2}^2}~.
\end{equation}

\noindent If the winds were accelerated near-instantaneously to their terminal speeds V$_{\infty}$ then
the wind parameters in Table 1 would give $r_{1}$ = 2.35 R$_{\odot}$ $<$ $R_{O9}$ so ram
balance could not be achieved and the higher momentum WR wind would shock onto the surface
of the O star.  But because of the close binary separation of CQ Cep the winds
are not expected to reach terminal speed along the line of centers. Thus it is
appropriate to use standard wind velocity profiles in equation (1) of the form
$V(r)$ = $V_{\infty}[1 - (R_{\rm star}/r)]^{\beta}$ where  $\beta$ $\approx$ $+$0.8 - $+$1.0
is typical  for hot stars (Lamers \& Cassinelli 1999).
Defining the dimensionless ratios 
P$_{WR/O}$ $\equiv$ $\dot{M}_{WR} V_{\infty,WR}$/$\dot{M}_{O9} V_{\infty,O9}$,
$x_{s} \equiv r_{1}/D$, and $C \equiv (R_{WR}/D) \approx (R_{O9}/D)$ 
for $R_{WR} \approx R_{O9}$ (De97) the above becomes

\begin{equation}
P_{WR/O} = \left(\frac{1}{x_{s}} - 1\right)^2\left[\left(\frac{1 - \frac{C}{x_{s}}}{1 - \frac{C}{1-x_{s}}}\right)^\beta + \frac{S}{P_{O/\nu}}\right]~ .
\end{equation}

\noindent The above equation is analogous to equation (4) of Gayley, Owocki, \& Cranmer (1997; hereafter GOC97)
except that we have not assumed the WR wind to have reached terminal speed at the stagnation point.
The last term on the right accounts for radiative braking of the WR wind by
the O star radiation field where $S$ is the reflection fraction  of the O-star radiative
momentum deposited into the WR wind near the interaction region. The ratio of wind to 
radiative momentum of the O star is  
$P_{O9/\nu}$ = ($\dot{M}_{O9} V_{\infty,O9} c)/L_{O9}$ = 0.46,
where $c$ is the speed of light and $L_{O9}$ = 1.16 $\times$ 10$^{5}$ L$_{\odot}$ (De97)
is the luminosity of the O star. The case $S$ = 0 corresponds to no braking and
reasonable limits on $S$ for the present case are 0.5 $\ltsimeq$ $S$ $\ltsimeq$ 2,
where the upper limit $S$ $\approx$ 2 corresponds to complete back reflection (GOC97).

Adopting $D$ = 20.4 R$_{\odot}$ and  R$_{\rm WR}$ $\approx$ R$_{\rm O9}$ = 8.2 R$_{\odot}$ 
yields $C$ $\approx$ 0.4 and the above equation must
be satisfied for 0.4 $<$ $x_{s}$ $<$ 0.6 to achieve ram balance between the stars. 
If braking is ignored ($S$ = 0) and  the right side of the above expression
is considered as a function $f(x_{s})$ versus $x_{s}$ then it is easily seen that
$f(x_{s} = 0.4)$ = 0, $f(x_{s} = 0.5)$ = 1, 
and $f(x_{s}) \rightarrow \infty$ as $x_{s} \rightarrow$ 0.6.
Thus, if the two stars had identical winds then ram balance would be achieved 
at the midpoint $x_{s}$ = 0.5. But for CQ Cep we have $P_{WR/O}$  = 58.5 (Table 1)
and ram balance can only be achieved  
at or near  the WR star surface ($x_ {s} \approx$ 0.599) where the WR wind is at 
effectively zero speed. This extreme condition  tells us that the O star wind
lacks sufficient momentum to achieve balance between the stars and again we
conclude that the WR wind overpowers the O star wind and the shock forms 
at or very near the O star surface ($x_{s} \approx$ 0.4).  If radiative braking
is included then the above conclusion does not change even for complete
back reflection ($S$ $\approx$ 2). 

In the above ram balance calculation, we have assumed a wind velocity law
of the form $V(r)$ = $V_{\infty}[1 - (R_{\rm star}/r)]^{\beta}$ and for
$R_{\rm star}$ we have used the photospheric radius. That is, we have
implicitly assumed that the winds of both stars are accelerated from
zero speed starting at the photosphere. This may be an oversimplification
for WR stars for which there is some observational evidence that the wind
has already reached appreciable speed at the photosphere (e.g. Schmutz 1997).
The precise form of the wind acceleration law in the optically thick part of
the wind below the photosphere is not well-known but the acceleration is 
thought to be slow ($\beta \sim$ 5; Nugis \& Lamers 2002). It is obvious
that if the WR wind is already at non-zero speed at the photosphere, then
the WR wind momentum between the stars is greater than we have assumed above,
strengthening the conclusion that the WR wind dominates the O star wind.

\subsubsection{Maximum Shock Temperature}

In its basic formulation, colliding wind shock theory predicts that the 
hottest plasma occurs  along the line-of-centers between the stars
near the stagnation point. For an adiabatic shock the  maximum plasma temperature
is predicted to be

\begin{equation}
kT_{cw,max} \approx 1.96\mu\left[\frac{V_{\perp}}{1000~ {\rm km~ s^{-1}}}\right]^{2}~{\rm keV},
\end{equation}

\noindent where $\mu$ is the mean mass (amu) per particle in the wind
and $V_{\perp}$ is the wind velocity component perpendicular
to the shock front (Fig. 4; Luo et al. 1990; Stevens et al. 1992).
The shocked plasma will span a range of temperatures with cooler
plasma present on the wings of the shock so the average temperature
as determined from X-ray observations will be less than the maximum
shock temperature given above. If the winds have reached terminal speeds
($V_{\infty}$ $\approx$  2000 km s$^{-1}$)
then maximum plasma temperatures of several keV are expected,
But if the winds have not reached terminal speeds before colliding,
as is likely the case for near-contact systems such as CQ Cep,
the maximum shock temperature  will be lower and the intrinsic
X-ray spectrum will thus peak at lower energies.

To be  more specific, we use the same wind velocity law as above 
$V_{WR}(r)$ = $V_{\infty}[1 - (R_{\rm WR}/r)]^{\beta}$
where $r$ is the distance from the center of the WR star and $\beta$ = $+$0.8.
For a binary separation $D$ = 20.4 R$_{\odot}$
and stellar radii $R_{\rm WR}$ = 8.2 R$_{\odot}$ and $R_{\rm O9}$ = 8.23 R$_{\odot}$ (De97), 
the velocity of the WR wind at the O-star 
surface ($r \approx$ 12.2 R$_{\odot}$) is $V(r)$ = 0.41$V_{\infty,{\rm WR}}$ $\approx$ 836 km s$^{-1}$
for $V_{\infty,{\rm WR}}$ = 2040 km s$^{-1}$. The above sub-terminal speed  
gives kT$_{cw,max}$ $\approx$ 1.8 keV for a He-dominated  WN star wind ($\mu$ = 4/3).
Because of its close binary separation,  CQ Cep is expected on theoretical
grounds to be in the non-adiabatic regime ($\chi$ $<<$ 1 in eq. [8] of St92),
and in that case radiative cooling will further soften the spectrum
and any radiative braking of the WN wind by the O star's radiation field
would also reduce V$_{WR}$. Taking these factors  qualitatively into account and 
recalling that the observed temperature is an average and not a maximum
value, we expect kT$_{cw,obs}$ $<$ kT$_{cw,max}$ $\approx$ 1.8 keV.

This above upper limit kT$_{cw,max}$ $\approx$ 1.8 keV should only be considered a 
rough estimate since it is sensitive to the assumed wind velocity profile, WN abundances, 
and stellar radii (Table 4 of De97), all of which are uncertain.    
Comparing the above upper limit 
with the hot plasma temperatures in Table 4, one can see that it is consistent with the lower
temperatures of models A and B but not with the higher temperatures of
models C and D. This discrepancy may not be significant in light of the uncertainties in the 
stellar parameters and wind velocity profile. As we have  noted above (Sec. 5.1.1) the 
WR wind may be slowly accelerated below the photosphere and speeds at the photosphere
of several hundred km s$^{-1}$ are possible (Nugis \& Lamers 2002). In that case the 
WR wind speed as it impacts the  O star surface would be greater than assumed above, leading to 
higher maximum shock temperatures.

\subsubsection{Colliding Wind X-ray Luminosity}

The predicted X-ray luminosity of a colliding wind shock impacting the 
O star surface in the radiative case is 
$L_{x,cw} = (\pi$/4)$\rho_{\infty}V_{\infty,WR}^{3}R_{\rm O9}^2$
(eq. [80] of Usov 1992), where $\rho_{\infty}$ is the wind mass density,
$V_{\infty,WR}$ is the terminal speed of the WR wind, and $R_{\rm O9}$ is
the O star radius. This can be rewritten as

\begin{equation}
L_{x,cw} =  \frac{1}{8}\left[\frac{R_{\rm O9}}{D}\right]^2L_{wind,WR}    
\end{equation}

\noindent where the terminal WR wind luminosity is
$L_{wind,WR}$ = (1/2)$\dot{M}_{WR}V_{\infty,WR}^2$ 
and we have used the expression for the wind  mass density  
$\rho_{\infty}$ =  $\dot{M}_ {WR}$/(4$\pi$$D^{2}$$V_{\infty,WR}$).
Equation (4) is valid  for  WR$+$O binaries where the 
WR wind has reached terminal speed  at the shock 
interface  and the separation $D$ between the centers  of the stars is 
much greater than the stellar radii. But in close binaries like CQ Cep
these conditions are not satisfied and equation (4) must be modified.
The relevant geometry for a close binary with the WR wind shocking onto the O star 
surface is shown in Figure 4. The distance $r$ from the center of the WR star
to the shock interface varies across the O star surface, as does the 
magnitude and direction of the WR wind velocity vector {\bf V}(r).
Also, at close separation the WR wind will in general not be at terminal speed 
upon impacting the O star. Taking these factors into account, 
the maximum colliding wind shock X-ray luminosity
is given by (Appendix A)

\begin{equation}
L_{x,cw} = \frac{1}{8}\left[\frac{R_{\rm O9}}{D}\right]^2 L_{wind,WR} \cdot  F
\end{equation}

\noindent where for CQ Cep we use $D$ = 20.4 R$_{\odot}$,
$R_{\rm O9}$/$D$ $\approx$ 0.4 (De97), and the WR star
terminal wind luminosity is $L_{wind,WR}$ = 
4.2 $\times$ 10$^{37}$ ergs s$^{-1}$ (Table 1). The factor $F$  is a 
correction term that takes into account the changes in the distance $r$ 
and the WR wind velocity vector across the shock interface (Fig. 4; Appendix A).
Evaluating $F$ numerically we obtain
$F$ =  0.224 and log $L_{x,cw}$ = 35.27  (ergs s$^{-1}$) for $\beta$ = +0.8.
Alternatively, we compute $F$ = 0.152 and log $L_{x,cw}$ = 35.11 (ergs s$^{-1}$)
for $\beta$ = +1.0. The above predicted values are about two orders of
magnitude greater than observed (Table 4).

It has been known for some time that the 
X-ray luminosities predicted by colliding wind theory for 
short-period WR binaries are about an
order-of-magnitude larger than observed. Cherepashchuk (1990) suggested
that the discrepancy could be accounted for if the wind is clumped.
If most of the mass is lost in the form of clumps then the effective
mass-loss rate of the unclumped (smooth) gas between the clumps that
interacts to form the colliding wind shock is lower than the total mass-loss
rate. In the Cherepashchuk (1990) picture, only $\approx$20\% of the 
total mass-loss rate is attributed to the smooth wind, in which case
the value in Table 1 would give a smooth (unclumped) mass-loss rate for
CQ  Cep of log $\dot{\rm M}_{\rm WR,smooth}$ $\approx$ $-$5.2
which is still too large to bring the predicted L$_{x}$ 
into agreement with the observation. Furthermore, a large downward correction of
$>$1.0 dex in the adopted WN star mass-loss rate (Table 1) seems very unlikely given the results of
the study of Nugis, Crowther, \& Willis (1998). They obtained 
log $\dot{\rm M}_{\rm clumpy}$ $-$ log $\dot{\rm M}_{\rm smooth}$ = $-$0.19 (1$\sigma$ = 0.28)
for WN stars. Nugis et al. applied some revisions to the  clumping-independent mass-loss 
estimates for CQ Cep based on  polarization techniques (St.-Louis et al. 1988) and found  
log $\dot{\rm M}_{\rm WR,pol}$ =  $-$4.46 (M$_{\odot}$ yr$^{-1}$).
It thus seems that for CQ Cep there is a significant 
discrepancy between the observed  L$_{x}$ and the larger values predicted by
colliding wind theory that is not currently explained by 
clumping-corrected WR  mass-loss  rates.

Another possible way to reduce the discrepancy between the predicted $L_{x,cw}$ 
and the observed $L_{x}$ is to assume a smaller radius for the companion star.
But if the companion is indeed an O-type star, as currently believed, then
a radius less than the value $R_{O9}$ = 8.23 R$_{\odot}$ assumed above is
quite unlikely given that almost all previous studies determine the companion
radius to be $\gtsimeq$9 R$_{\odot}$  (Table 4 of De97). 
One exception is the study of Underhill et al. (1990) which concluded
that the companion could be a B0 star with radius $R_{B0}$ = 6.6 R$_{\odot}$, 
but the validity of their  model has been questioned by De97. Even if the companion  radius 
were as small as 6.6 R$_{\odot}$ it would only reduce the $L_{x,cw}$ prediction (eq. [4])
by 0.2 dex. Thus, even a companion radius at the low end of the estimated range
corresponding to a B0 star is not sufficient to resolve the X-ray luminosity
discrepancy with colliding wind predictions.

\subsection{Orbital Effects}

In the framework of the colliding wind shock picture, the 
X-ray emission is expected to change as the O star passes
in front of the WR star. As discussed above (Sec. 5.1.1), the stagnation
point will lie at or near the surface of the O star and
plasma at maximum shock temperatures kT$_{cw,max}$ $\approx$ 1.8 keV
should be present along the line-of-centers. Since the CQ Cep orbit
is viewed at high inclination (i.e. nearly edge-on), any hot
plasma at the stagnation point along the line-of-centers will  be partially occulted when
the O star  is in front because in that viewing geometry
the stagnation point lies behind the  O star  and is
obstructed from full view. Lower X-ray absorption is
also anticipated when the O star is  nearly in front {\em if} 
the X-rays originate between the stars because 
the emission is then viewed through the O star wind, which
is less dense than the WR wind.
Shortly after the O star  passes in front and the system
moves toward quadrature, any hot colliding wind shock
plasma located near the line-of-centers should be
revealed along with an increase in observed hard-band flux (or count rate).

But our comparison of the count rates and observed fluxes 
between the first half of the observation when the O-star
was in front and the second half show that any
effect on the X-ray emission by the eclipse at $\phi$ = 0 was quite
small, with the observed (absorbed) flux in the second half being
at most 4\% greater than the first half. 
Also, as noted (Sec. 4.2) the median photon energy and hardness ratios in the
first and second halves of the observation are nearly identical.
Furthermore, our fits of separate  spectra extracted for the 
first and second halves show no significant change in the 
column density associated with the cool X-ray component.
Since wind absorption affects mainly lower energy photons, it
is the cool component that would be most affected by any
changes in column density toward the source.

Since the expected orbital variability was not detected we must
conclude  that any existing hot colliding wind plasma 
is not strongly localized along the line-of-centers. 
It is very likely  that the geometry of the wind interaction region 
in CQ Cep is affected by the close binary spacing and orbital motion. 
On the basis of asymmetric optical eclipse light curves, De97 concluded
that the system is in an overcontact configuration but 
Harries \& Hilditch (1997) found that their light curve solutions
required only  marginal contact. In either case, the region between
the stars will be tightly confined and this will affect wind  launching 
and acceleration. If wind development between the stars is 
impeded then there may be no  hot colliding wind shock plasma on
or near the line-of-centers. However, a colliding wind shock may still
be able to form away from the line-of-centers where the WR
wind is able to accelerate as it flows around the O star 
(see Fig. 3 of Luo et al. 1990). In this oblique geometry  the 
maximum predicted shock temperature will be less than for a head-on wind 
collision  because the WR wind velocity vector is no longer perpendicular 
to the shock interface  and the 
projected value $V_{\perp}$ in eq. (3) is reduced. 
Since the predicted value of kT$_{cw,max}$ $\approx$ 1.8 keV for a head-on
collision is already at or below the observed hot plasma temperature (kT$_{2}$), 
any further downward correction due to oblique geometry would only worsen 
the agreement.

Orbital motion  will also modify the shape of the wind 
interaction region in CQ Cep and will affect wind energetics
as a result of enhanced turbulence, as has been shown in hydrodynamic
simulations of the similar close binary system V444 Cyg (WN5$+$O6; P$_{orb}$ = 4.2 d).
The simulations of Walder (1995) show that the wind interaction
zone assumes a spiral shape and is shifted in the direction opposite
to orbital motion. The orbital speed of the 
stars in CQ Cep is $~\sim$320 km s$^{-1}$ assuming that the 
stars have similar masses (De97). This is considerably less
than the terminal wind speeds ($\sim$2000 km s$^{-1}$) but is 
similar to  that adopted for the WR star in V444 Cyg by Walder (1995).

Detailed hydrodynamic simulations are needed  to 
model the geometry of the wind interaction region in CQ Cep.
The development of realistic simulations will be challenging because
they will need to take into account wind development and acceleration
in the confined region between the  stars, radiative cooling, and orbital motion. 
A crucial question to be answered by future simulations is whether hot 
colliding wind plasma at the observed {\em Chandra} temperatures
of $\sim$2 - 3 keV can form in this close system and, if so, 
is its spatial distribution consistent with the very low levels of
X-ray variability (if any) seen in the {\em Chandra} data during optical
eclipse.

\subsection{Comparison With Other Close WN Binaries}

The absence of significant  X-ray variability in CQ Cep over half an orbit
is unusual for a WR$+$O binary but at least one other close
WR binary shows similar behavior. The WN3 $+$ OB system WR 46
has a complex optical light curve which shows multiple photometric  periodicities of
$\approx$0.14 - 0.3 d as well as  radial velocity variations of 0.329 d (7.9 hr).
The variability may be related to binarity as discussed in detail by 
Gosset et al. (2011) but other possible causes such as non-radial pulsations 
have also been proposed (H\'{e}nault-Brunet et al. 2011).

A 70 ks {\em XMM-Newton} observation of WR 46 was analyzed by Gosset et al. (2011),  
covering two cycles of the 0.329 d period.   They detected both cool and hot plasma
in the X-ray spectrum at temperatures similar to those found here for CQ Cep,
but an additional very soft component at kT $\approx$ 0.2 keV may also be present.
Based on statistical timing analysis in several different energy bands, 
they concluded that X-ray variability was possibly present but only in the 
lowest energy band (0.2 - 0.5 keV) and of low significance. The {\em XMM-Newton}
data were also analyzed by H\'{e}nault-Brunet et al. (2011). Their broad-band
(0.3 - 10 keV) EPIC $pn$ light curve shows low-amplitude fluctuations at the 
$\pm$20\% level but no results of  statistical variability tests that would 
substantiate real variability and quantify its significance were given. 
Their X-ray periodogram shows a peak at a period  P = 9.0 $\pm$ 2.5 hr and 
a similar period may be present in UV light curves. 
H\'{e}nault-Brunet et al. argued that non-radial pulsations may be
responsible for the low-level variability. However,  a further examination of 
the WR 46 {\em XMM-Newton} data  was carried out by Zhekov (2012) who applied 
$\chi^2$ analysis and concluded that the X-ray light curves are consistent with 
a constant count-rate source.  Based on the above studies, it is clear that
any X-ray variability  present in the WR 46 {\em XMM-Newton} data is of low 
significance and even  if short-period ($\sim$8 - 9 hr) modulations exist
they may not be due to binarity. The presence of hot plasma with little 
or no significant X-ray variability is reminiscent of our findings for
CQ Cep and suggests that the hot X-ray plasma may not arise solely in a
colliding wind shock, at least if the assumption that the hot plasma 
originates entirely in the close  WN$+$OB binary itself is correct.

To explain the {\em hot} X-ray plasma in WR 46, Gosset et al. suggested  
that a third more distant companion may be present and is orbiting
the central WN3 $+$ OB system with a period longer than a few weeks.
If the third object were a massive star, then the interaction of its
wind with the aggregate wind of the close binary could give rise to colliding wind shock
emission for which orbital X-ray variability on timescales of less 
than a  few weeks would not be expected. This scenario could
account for the lack of significant X-ray variability of WR 46
on $\sim$1-day timescales as well as the presence of hotter plasma  because
the winds would be at or near terminal speeds when colliding.
Similarly, a hypothetical third star in a wider orbit could be invoked 
to explain the hot plasma and lack of short-term ($\sim$1 day)
X-ray variability in CQ Cep. Gaposchkin (1944) speculated that a
third body could be responsible for perceived small changes in the  
time between primary and secondary minima in the optical light curve
of CQ Cep.  However,  an attempt by Borkovits \& Heged\"{u}s (1995) to find a  good 
third-body orbit for CQ Cep based on historical optical data
was not successful and no optical nebulosity was detected around CQ Cep in 
the study of Miller \& Chu (1993).

Until compelling evidence for a third star orbiting CQ Cep 
is presented,  the possibility of intrinsic hot X-ray plasma from 
one or both stars in the near-contact WN$+$OB system originating by a process
other than colliding winds should not be dismissed.
A point not to be overlooked here is that
X-ray luminosities of WN stars for which no evidence of binarity has
yet been found lie in the range log L$_{x}$ = 31.14 (WR 16) -
33.57 (WR 20b) ergs s$^{-1}$ (Skinner et al. 2010; 2012).
This range easily encompasses the luminosities of both 
WR 46 (log L$_{x}$ = 32.77; Zhekov 2012) and CQ Cep (Table 4).
It is of course possible that some of the putatively single WN
stars detected  previously in X-rays are close binaries with
unseen companions, but until that is demonstrated it would be
premature to discount intrinsic emission from the WN star or its
wind as a non-negligible contributor to the  X-ray emission detected from 
CQ Cep.  

\subsection{Magnetically-Confined Winds}

The current state of knowledge provides very few theoretical alternatives
to colliding winds for explaining the existence of hot plasma (kT $\gtsimeq$ 2 keV)
in  WR$+$O binaries. One plausible  alternative is based
on the magnetically-confined wind shock (MCWS) model that 
has been used to explain hot plasma in some O stars such
as the O7 V star $\Theta^{1}$ Ori C (Babel \& Montmerle 1997; hereafter BM97).
In this picture, sufficiently strong stellar magnetic fields
can confine the ionized wind and channel it toward the magnetic
equator where the oppositely-directed streams collide to
form a MCWS. This mechanism could be operating in the O9
component of CQ Cep but very strong B fields would be 
needed to confine the much stronger wind of the WN star.

The magnetic field strength needed to confine
the O-star wind can be estimated using the confinement parameter
$\eta$ = B$_{eq}^2$R$_{*}^2$/$\dot{M}$v$_{\infty}$
where B$_{eq}$ is the field strength at the magnetic
equator (ud-Doula \& Owocki 2002). 
For $\eta$ $\approx$ 1 the wind is  marginally-confined by the 
B-field. Adopting a radius R$_{*}$ = R$_{\rm O9}$ = 8.2 R$_{\odot}$
for the O star in CQ Cep and using the wind parameters
in Table 1, marginal wind confinement requires 
B$_{eq}$ $\sim$ 150 G and larger fields are needed for
strong confinement. A $\sim$150 G field would be sufficient
to account for the observed X-ray luminosity (eq. [10] of
BM97) and the O9 terminal wind speed
could produce MCWS plasma temperatures of several keV
(eq. [4] of BM97).

Since the MCWS is predicted to form near the magnetic 
equator of the O star, the hot plasma should be at
least partially occulted as the WN star passes in 
front ($\phi$ = 0.5), with an accompanying  drop in the 
hard-band count rate. There is some suggestion of a 
small decrease in the hard-band count rate near 
$\phi$ $\approx$ 0.5 in Fig. 2-bottom but the change is of low
significance ($\approx$1.5$\sigma$) and we lack phase coverage
beyond $\phi$ = 0.52. 

If the O9 star is producing hard X-rays in a MCWS then
rotational X-ray modulation may be present, as is the 
case for $\Theta^{1}$ Ori C which shows periodic modulations
at its 15.4 d rotation period.
Longer time monitoring of CQ Cep would be needed to determine
if periodic X-ray variability is present, but  none
is obvious in our short $\sim$1-day X-ray light curve.
One argument against the scenario in which most of the 
X-rays originate in the O9 star via a MCWS is the 
strong resemblance of the {\em Chandra} X-ray spectrum
of CQ Cep to that of the presumably single WN6 star WR 134 (Fig. 3).  
Similarly, Stickland et al. (1984) have noted that the
{\em IUE} spectrum of CQ Cep is ``virtually indistinguishable
from that of single WN7 stars and shows no clear evidence of
any OB companion spectrum''.
The above similarities  suggest that it is the WN star and not
the O-type companion that plays the dominant role in the 
X-ray (and UV)  emission of CQ Cep.

\subsection{Orbital Period of CQ Cep}

The time of primary optical minimum ($\phi$ = 0) 
predicted by the results  of De97 is 3.26 ks 
{\em earlier} than the minimum observed by {\em Chandra}
(Fig. 2-top). This difference is too large to explain on the basis 
of the formal uncertainties in the published De97 elements.
If we instead use the time-of-minimum-light equation of 
Walker et al. (1983; hereafter Wa83) then the predicted primary minimum is at
JD 2415000.410 $+$ (1.6412436 $\pm$ 0.0000009 d)$\cdot$E,
which occurs 2.84 ks {\em later} than the  {\em Chandra}
minimum. Again, the difference cannot be explained by the
formal uncertainty in the Wa83 period which accumulates to
1.96 ks at the epoch of our {\em Chandra} observation. Interestingly,
Harries \& Hilditch (1997) found a similar difference
of 1.56 ks and in the same sense (i.e. their observed time is
earlier than predicted by Wa83).

One possible explanation for the above discrepancies is that
the orbital period of CQ Cep is variable. The question of  orbital period variability
has been investigated in previous studies but the results are
contradictory. Wa83 found that 68 years of optical data 
analyzed in their study were consistent with no period change
greater than an annual  fractional change of $\sim$5 $\times$ 10$^{-7}$.
In contrast, the study of 
Antokhina et al. (1982; hereafter A82) found that the period was becoming
shorter at a rate of $-$0.019 ($\pm$0.006) s yr$^{-1}$ adding
support to a previous claim for a period change by Gaposchkin (1944).

As the conflicting results summarized above show, the presence of any orbital
period variability in CQ Cep remains  an open question.
Wa83 pointed out that other factors besides period variability 
such as eclipse asymmetries  could be contributing to
discrepancies in reported periods. The period change found by A82
is much too small to explain the difference between the 
time of {\em Chandra} optical minimum and that predicted
by the De97 results, and in any case the period change is 
of opposite sign needed to resolve the discrepancy. But, a 
small error in the value of the adopted orbital period 
from De97 could easily explain the difference, as shown below.

Three of the most recent P$_{\rm orb}$ determinations for
CQ Cep are 1.64122299 $\pm$ 2e-07 d (De97), 
1.6412436 $\pm$ 9e-07 d (Wa83), and
1.641249 d (A82; no error given).
The average of these three values is
$\overline{{\rm P}}_{orb}$ = 1.6412408 ($+$8.2e-06,$-$1.1e-05) d
where the uncertainties simply reflect the range in published
values. It is clear that the range  is
much larger than the formal uncertainties given  by
De97 and Wa83. If the above value of $\overline{{\rm P}}_{orb}$
is used in the time-of-minimum-light equation of De97 (Sec. 4.2),
then the primary minimum is predicted to occur
at t = 14.746 ks after the start of the {\em Chandra}
observation, which is in much better agreement with the 
observed minimum at t = 14.46 $\pm$ 0.02 ks. If instead
we start with the observed time of  {\em Chandra} minimum light
(HJD = 2456371.20268 $\pm$ 0.00023 [UTC]) and adopt the reference date
for primary minimum of De97 (HJD = 2450267.43158 $\pm$ 0.00077)
then the time difference divided by 3719 full orbits gives
P$_{\rm orb,CXO}$ = 1.6412400 $\pm$ 1.5e-7 d, where the 
error takes into account the uncertainty in the De97
epoch date and in the time of  {\em Chandra} minimum light.
This period  differs by only 9e-07 d from the average value 
given above. Thus, the difference between the  time 
of optical minimum observed by {\em Chandra} and that
predicted by the results of De97 can be explained by
adopting an orbital period P$_{\rm orb,CXO}$ = 1.6412400 $\pm$ 1.5e-7 d,
which is slightly larger than the value determined by De97 but
in quite good agreement with the average orbital period determined
from three previous studies.

Further optical monitoring of CQ Cep would be worthwhile
to confirm the result of A82 that the period is slowly decreasing
and, if so, to determine if the  rate-of-change is stable. A reliable 
determination of the rate-of-change in the period can be
used to constrain the mass-loss rate and the rate at which mass is flowing
from the WN star onto the O star, as shown by A82
for CQ Cep and by Khaliullin (1974) for the short-period
WN$+$O binary V444 Cyg.

\section{Summary}

The main results of this study are the following:

\begin{enumerate}

\item {\em Chandra} has detected luminous X-ray emission from 
CQ Cep. A simultaneous  {\em Chandra} optical light curve 
clearly shows a deep  eclipse when
the O star passed in front of the WN star.  
However, little or no X-ray variability was seen during 
the observation. 

\item The {\em Chandra} ACIS-S spectrum shows numerous
blended emission lines and is quite similar to spectra
of apparently single WN6 stars such as WR 134. The spectrum can be fitted with either 
2T optically thin plasma models or 2T plane-parallel shock models.
Such models give a cool component plasma temperature kT$_{1}$ $\approx$ 0.6 keV
and a hot component temperature kT$_{2}$ $\approx$ 2 - 3 keV, similar
to temperatures observed in single WN stars and other close WN$+$O binaries.

\item The unabsorbed X-ray luminosity of CQ Cep log L$_{x}$ = 33.25 - 33.48 ergs s$^{-1}$
is at the high end of the range observed for WR stars. This value is similar to that of
other close WN$+$O binaries but is also comparable to that of the most luminous
single (non-binary) WN stars.

\item The absence of significant X-ray variability during optical eclipse of
this high-inclination system is contrary to expectations if the hottest plasma 
originates in a colliding wind shock and is localized along the line-of-centers 
between the stars. The predicted temperature of a colliding wind shock 
formed by the WR wind impacting the O star at sub-terminal speed is 
marginally consistent with the lower values determined from X-ray spectral fits
but the  predicted X-ray luminosity is at least an order of magnitude higher
than observed. Thus, we find discrepancies with colliding wind model 
predictions that remain to be explained.

\item The lack of significant X-ray variability in CQ Cep during optical
eclipse at $\phi$ = 0 (O star in front) provides compelling evidence that 
the X-ray plasma is extended on scales comparable to or larger than the 
binary system (binary separation $\sim$20 R$_{\odot}$). Radiative shocks 
distributed far out in the winds of one or both stars remain a plausible 
explanation for the cool plasma in CQ Cep. However detailed hydrodynamic
simulations of the wind interaction region that take into account the  
wind velocity field between the stars, radiative cooling, high inclination viewing 
geometry, and orbital motion effects will be needed to determine whether the 
presence of hot plasma (kT $\approx$ 2 - 3 keV) with little or no X-ray 
variability over half an orbit is consistent with colliding wind theory.

\item Using the time of primary optical minimum in the {\em Chandra} ACA
light curve and the reference date of primary optical minimum
determined by De97, we obtain an orbital period for CQ Cep 
of ${\rm P_{orb}}$ = 1.6412400 $\pm$ 1.5e-7 d, in good agreement
with the average period computed from three previous optical studies.

\end{enumerate}  

\acknowledgments

This work was supported by {\em Chandra} award GO3-14006X
issued by the Chandra X-ray Observatory Center (CXC). The CXC is operated by the
Smithsonian Astrophysical Observatory (SAO) for, and on behalf of,
the National Aeronautics Space Administration under contract NAS8-03060.

\newpage

\appendix
\section{Colliding Wind X-ray Luminosity for Close WR+O Binaries}

We derive here an expression for the colliding wind X-ray luminosity
in close WR$+$O binaries such as CQ Cep where the stellar radii
$R_{WR}$ and R$_{O}$ are not much less than the distance $D$
between the centers of the two stars. In this case the magnitude and
direction of the WR wind velocity vector change across the shock 
interface at the O star surface.

We assume a radial spherically-symmetric WR wind which overpowers
the O star wind so that a colliding wind shock forms at the O
star surface. As shown in Figure 4, the WR wind velocity vector
${\bf V}(r)$ impacts the O star surface at a distance $r$ from 
the center of the WR star. The velocity component perpendicular
to the shock interface is $V_{\perp}(r)$. We assume a standard wind velocity
profile $V(r) = V_{\infty}[1 - (R_{WR}/r]^{\beta}$ where
$V_{\infty}$ is the terminal speed of the WR wind. The wind mass density
is $\rho(r)$. As Figure 4 shows, a point on the O star surface at 
distance $r$ corresponds to a specific value of  $\alpha$, where
$\alpha$ is the angle formed  at the O star's center between the line-of-centers
and the point on the O star surface. The dependence of $r$ on $\alpha$ is given 
by the law of cosines as $r^2$ = $D^2 + R_{O}^2 - 2DR_{O}{\rm cos}\alpha$. 
Adopting   $\alpha$ as the independent
variable, the maximum X-ray luminosity is obtained by integrating the
energy flux crossing the shock over the portion of the O star
surface that is intercepted by the WR wind, 

\begin{equation}
L_{x,cw} = \frac{1}{2}\int_{0}^{\alpha_{max}}\rho(\alpha)V_{\perp}(\alpha)^3\cdot R_{O}^2 2\pi {\rm sin}\alpha~ d\alpha~ .
\end{equation}

\noindent The limits on the integral are from $\alpha$ = 0 (the stagnation point)
to $\alpha_{max}$ at the tangent point, where cos $\alpha_{max}$ = $R_{O}/D$. The WR wind mass density
can be written as

\begin{equation}
\rho(\alpha) = \frac{\dot{M}_{WR}}{4\pi r(\alpha)^2V(\alpha)}~.
\end{equation}

\noindent We now define the dimensionless quantities $\xi_{O}$ $\equiv$ $R_{O}$/$D$,
$\xi_{WR}$ $\equiv$ $R_{WR}$/$D$ and the functions

\begin{equation}
f(\alpha) = 1 - 2\xi_{O}{\rm cos}\alpha + \xi_{O}^2
\end{equation}

\begin{equation}
g(\alpha) = \left(1 - \frac{\xi_{WR}}{\sqrt{1 - 2\xi_{O}{\rm cos}\alpha + \xi_{O}^2}}\right)~.
\end{equation}

\noindent From Figure 4 it can then be seen that 

\begin{equation}
V(\alpha) = g(\alpha)^{\beta}V_{\infty}
\end{equation}

\begin{equation}
V_{\perp}(\alpha) = g(\alpha)^{\beta}{\rm cos}(\alpha + \omega)V_{\infty} ~.
\end{equation}

\noindent Inserting eq. (A2) into eq. (A1) and then using (A3) - (A6) gives

\begin{equation}
L_{x,cw} = \frac{1}{8}\left(\frac{R_{O}}{D}\right)^2 L_{wind,WR} \cdot \left[4\int_{0}^{\alpha_{max}}\frac{g(\alpha)^{2\beta}}{f(\alpha)}{\rm cos}^3(\alpha + \omega) {\rm sin}\alpha~d\alpha\right]
\end{equation}

\noindent where the WR terminal wind luminosity is $L_{wind,WR}$ = (1/2)$\dot{M}_{WR}V_{\infty,WR}^2$
and cos $\alpha_{max}$ = $R_{O}/D$. The 
quantity $F$ in square brackets in eq. (A7) can be computed numerically, noting that 
$\omega$ is uniquely determined by $\alpha$.

\newpage

                                                                                                                       
\begin{deluxetable}{lccccccc}
\tabletypesize{\scriptsize}
\tablewidth{0pt}
\tablecaption{Stellar Properties of CQ Cep}
\tablehead{
           \colhead{Type}               &
           \colhead{dist.}              &
           \colhead{A$_{V}$}            &
           \colhead{V$_{\infty,wr}$}            &
           \colhead{log $\dot{\rm M}_{wr}$}            &
           \colhead{V$_{\infty,o9}$}         &
           \colhead{log $\dot{\rm M}_{o9}$}            &
           \colhead{P$_{orb}$} \\            
           \colhead{}                   &
           \colhead{(kpc)}                   &
           \colhead{(mag)}                   &
           \colhead{(km s$^{-1}$)}         &
           \colhead{(M$_{\odot}$ yr$^{-1}$)}         &
           \colhead{(km s$^{-1}$)}                   &
           \colhead{(M$_{\odot}$ yr$^{-1}$)}                   &
           \colhead{(d)}                   
}
\startdata
WN6$+$O9II-Ib  & $\sim$2.75 - 3.5  &  2.0 - 2.2           & 2040  & $-$4.5  & $\sim$2200                 & $\sim$ $-$6.3            & 1.6412   \\
(1,2)          & (2,3,4)           &  (2,3,4)             & (3,5) & (3,5)   & (6,7)                      & (6,7)                    & (2)     \\
\enddata
\tablecomments{Spectral type is from Marchenko et al. (1995) but Demircan et al. (1997)
have noted that the companion may be a O9 III or O9 V star.  The O9 wind parameters are  
typical values for O9II-I.
Published values of the orbital period show small but significant differences (Sec. 5.5). \\
References:~ (1) Marchenko et al. (1995)~(2) Demircan et al. (1997)~(3) Stickland et al. (1984)
~(4) van der Hucht (2001) ~(5) Hamann et al. (2006) ~(6) Howarth \& Prinja (1989) ~(7) Prinja et al. (1990).
}
\end{deluxetable}

                                                                                                                                             
\begin{deluxetable}{llcccccl}
\tabletypesize{\scriptsize}
\tablewidth{0pt}
\tablecaption{ X-ray Properties of CQ Cep (Chandra ACIS-S)}
\tablehead{
           \colhead{R.A.}               &
           \colhead{decl.}              &
           \colhead{Net Counts}         &
           \colhead{H.R.}               &
           \colhead{E$_{50}$}           &
           \colhead{P$_{const}$}             &
           \colhead{F$_{x}$}             &
           \colhead{Optical Position}      \\
           \colhead{(J2000)}                 &
           \colhead{(J2000)} &
           \colhead{(cts)}                                          &
           \colhead{}                                          &
           \colhead{(keV)}                                          &
           \colhead{}                                          &
           \colhead{(ergs cm$^{-2}$ s$^{-1}$)}                                          &
           \colhead{(arcsec)}
                                  }
\startdata
 22 36 53.95 & $+$56 54 20.93 & 2086 $\pm$ 46  & $-$0.278 & 1.77 & 0.77     & 2.03 $\times$ 10$^{-13}$ & J223653.955$+$565420.98 (0.06) \\
\enddata
\tablecomments{
The nominal pointing position for
the observation was (J2000.0) RA = 22$^h$ 56$^m$ 54.11$^s$,
decl. = $+$56$^{\circ}$ 54$'$ 31$''$.7.
X-ray data are from CCD7 (ACIS chip S3) using events in the 0.2 - 8 keV range inside a
circular source extraction region of radius 1$''$.8 centered on the X-ray peak.
Tabulated quantities are: J2000.0 X-ray centroid position (R.A., decl.), total source counts accumulated
in a 79,391 s live time exposure,  hardness ratio H.R. = (H$-$S)/(H$+$S) where H = counts(2-8 keV)
and S = counts(0.2 - 2 keV), median photon  energy (E$_{50}$),
probability of constant count-rate determined by the Gregory-Loredo
algorithm (P$_{const}$); absorbed X-ray flux in the  0.3 - 8 keV band (F$_{x}$;  see  Table 4 for
model-dependent values),
and {\em HST} Guide Star Catalog counterpart J2000 optical position.
The offset in arcseconds  between the X-ray and optical position is given in parentheses.}

\end{deluxetable}

\begin{deluxetable}{llll}
\tabletypesize{\scriptsize}
\tablewidth{0pt}
\tablecaption{CQ Cep X-ray Count Rates (Chandra ACIS-S)}
\tablehead{
           \colhead{Time Range}               &
           \colhead{          }           &
           \colhead{Energy    }           &
           \colhead{              }       \\
           \colhead{          }               &
           \colhead{0.2 - 8 keV    }           &
           \colhead{0.2 - 2 keV    }           &
           \colhead{2   - 8 keV    }                   
}
\startdata
all      & 25.0 (4.3)  & 16.0 (3.4) & 9.1 (2.1) \nl
1st half & 23.8 (3.9)  & 15.1 (3.1) & 8.8 (2.1) \nl
2nd half & 26.2 (4.0)  & 16.8 (3.1) & 9.4 (2.4) \nl
\enddata
\tablecomments{Rates were computed from light curves binned at 2000 s intervals and  are in units of c/ks.
               The standard deviation is given in parentheses.}
\end{deluxetable}

\newpage


\begin{deluxetable}{lllll}
\tabletypesize{\scriptsize}
\tablewidth{0pc}
\tablecaption{{\em Chandra} Spectral Fits of CQ Cep
   \label{tbl-1}}
\tablehead{
\colhead{Parameter}      &
\colhead{ }              &
\colhead{  }
}
\startdata
Model                                              & A                          & B                                  & C                                   & D   \nl
Type\tablenotemark{a}                              & 2T $vapec$\tablenotemark{a}& 2T $vapec$\tablenotemark{a}        & 2T $vapec$\tablenotemark{a}         & 2T $vpshock$\tablenotemark{a}  \nl
Abundances                                         & solar                      & nonsolar\tablenotemark{b}          & solar$+$nonsolar\tablenotemark{c}   & solar$+$nonsolar\tablenotemark{c} \nl
N$_{\rm H,ISM}$ (10$^{21}$ cm$^{-2}$)              & \{4.4\}\tablenotemark{d}   & \{4.4\}\tablenotemark{d}           & \{4.4\}\tablenotemark{d}            &  \{4.4\}\tablenotemark{d}       \nl
N$_{\rm H,1}$ (10$^{21}$ cm$^{-2}$)                & 7.59 [6.60 - 8.70]         & 0.07 [0.06 - 0.08]\tablenotemark{e}& 6.92 [5.70 - 8.10]                  & 4.45 [3.06 - 6.60]   \nl
N$_{\rm H,2}$ (10$^{21}$ cm$^{-2}$)                & 25.1 [19.4 - 35.6]         & 0.37 [0.31 - 0.47]\tablenotemark{e}& 8.97 [6.10 - 13.0]\tablenotemark{e} & 0.25 [0.21 - 0.33]\tablenotemark{e}  \nl
kT$_{1}$ (keV)                                     & 0.57 [0.49 - 0.62]         & 0.56 [0.47 - 0.61]                 & 0.58 [0.53 - 0.63]                  & 0.63 [0.49 - 0.72]  \nl
kT$_{2}$ (keV)                                     & 1.82 [1.48 - 2.08]         & 1.96 [1.70 - 2.27]                 & 2.34 [1.90 - 2.87]                  & 3.30 [2.68 - 3.87]  \nl
norm$_{1}$ (10$^{-4}$ cm$^{-5}$)\tablenotemark{f}  & 3.76 [2.86 - 5.32]         & 3.30 [2.46 - 4.15]                 & 3.12 [2.22 - 4.25]                  & 1.03 [0.60 - 2.22]   \nl
norm$_{2}$ (10$^{-4}$ cm$^{-5}$)\tablenotemark{f}  & 4.50 [3.64 - 6.63]         & 2.86 [2.29 - 3.42]                 & 0.85 [0.65 - 1.18]                  & 2.27 [1.99 - 2.85]  \nl
$\tau_{1}$ (10$^{11}$ cm$^{-3}$ s)                 & ...                        & ...                                & ...                                 & 2.78 [0.24 - 14.0] \nl
$\tau_{2}$ (10$^{11}$ cm$^{-3}$ s)                 & ...                        & ...                                & ...                                 & 2.14 [1.40 - 8.40] \nl
$\chi^2$/dof                                       & 89.5/76                    & 96.3/76                            & 90.3/76                             & 71.1/74             \nl
$\chi^2_{red}$                                     & 1.18                       & 1.27                               & 1.19                                & 0.96               \nl
F$_{\rm X}$ (10$^{-13}$ ergs cm$^{-2}$ s$^{-1}$)   & 2.04 (16.7)                & 2.06 (13.0)                        & 2.00 (12.3)                         & 2.09 (20.6)        \nl
F$_{\rm X,2}$ (10$^{-13}$ ergs cm$^{-2}$ s$^{-1}$) & 1.41 (6.80)                & 1.45 (6.32)                        & 1.41 (4.14)                         & 1.74 (14.4)       \nl
log L$_{\rm X}$ (ergs s$^{-1}$)                    & 33.39                      & 33.28                              & 33.25                               & 33.48             \nl
\enddata

\tablecomments{
Based on  XSPEC (vers. 12.8.2) fits of the background-subtracted ACIS-S spectra binned
to a minimum of 20 counts per bin using 79,391  s of  exposure time. The spectra were
modeled using  an absorbed  two-temperature (2T)  optically thin plasma model
or plane-parallel shock model.
The absorption of each plasma component was allowed to vary independently,
but the source and absorber abundances for each component were forced to be the same.
The tabulated parameters
are ISM absorption column density (N$_{\rm H,ISM}$; solar abundances),
wind absorption column density (N$_{\rm H,i}$), plasma energy (kT),
XSPEC component normalization (norm), and upper limit on the ionization timescale ($\tau$) for the shock model.
Solar abundances are referenced to the  values of Anders \& Grevesse (1989).
WN abundances are referenced to the generic values given by  van der Hucht et al. (1986).
Square brackets enclose 90\% confidence intervals.
Quantities enclosed in curly braces were held fixed during fitting.
The total X-ray flux (F$_{\rm X}$) and the flux associated with the high-temperature component
(F$_{\rm X,2}$)  are the absorbed values in the 0.3 - 8 keV range, followed in
parentheses by  unabsorbed values. The unabsorbed values were measured by
setting all N$_{\rm H}$ values in the model to zero.
The total X-ray luminosity L$_{\rm X}$  is the  unabsorbed
value in the 0.3 - 8 keV range and  assumes a distance d = 3.5 kpc. }

\tablenotetext{a}{Model is of form N$_{\rm H,ism}$$\cdot$N$_{\rm H,1}$$\cdot$kT$_{1}$ $+$
                                      N$_{\rm H,ism}$$\cdot$N$_{\rm H,2}$$\cdot$kT$_{2}$ .
                  In XSPEC, the absorption column density N$_{\rm H}$ was modeled using
                  the $wabs$ model (solar abundances) and the $vphabs$ model (nonsolar abundances).
                  The optically thin plasma components kT were modeled using the $vapec$ model.
                  The plane-parallel shock components kT in model D were modeled using the $vpshock$ model
                  with version 2.0 non-equilibrium ionization data ($neivers 2.0$). }
\tablenotetext{b}{Held fixed at the generic values for WN stars given in Table 1 of van der Hucht et al. (1986).
The WN abundances reflect H depletion and N enrichment and
are by number:
He/H = 14.9, C/H = 1.90E-03, N/H = 9.36E-02,
O/H = 4.35E-03, Ne/H = 9.78E-03, Mg/H = 3.26E-03,
Si/H = 3.22E-03, P/H = 1.57E-05, S/H = 7.60E-04,
Fe/H = 1.90E-03. All other elements were held fixed
at solar abundances (Anders \& Grevesse 1989). }
\tablenotetext{c}{Cool plasma component fixed at solar abundances and hot component fixed at generic
                  WN abundances.}
\tablenotetext{d}{Held fixed during fitting at the value inferred from A$_{\rm V}$ = 2 mag
                  and the conversion  N$_{\rm H}$ $\approx$ 2e21 $\cdot$ A$_{\rm V}$ (Gorenstein 1975).
                  Solar abundances assumed for ISM absorption.}
\tablenotetext{e}{Absorption is He-dominated for WN abundances.}
\tablenotetext{f}{For thermal $vapec$ models, the norm is related to the volume emission measure
                  (EM = n$_{e}^{2}$V)  by
                  EM = 4$\pi$10$^{14}$d$_{cm}^2$$\times$norm, where d$_{cm}$ is the stellar
                  distance in cm. At d = 3.5 kpc this becomes
                  EM = 1.47$\times$10$^{59}$ $\times$ norm (cm$^{-3}$). }
\end{deluxetable}



\begin{figure}
\figurenum{1}
\epsscale{1.0}
\includegraphics*[width=10.0cm,height=12.5cm,angle=0]{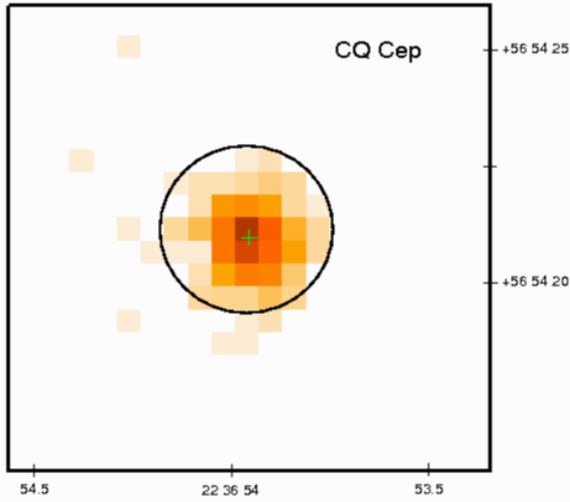}
\caption{Broad-band (0.2 - 8 keV)  ACIS-S image of CQ Cep.
The circle of  radius 1.$''$8 shows the region used to extract
the X-ray source spectrum and light curves.
The $+$ symbol marks the {\em HST} Guide Star Catalog optical position.
Pixel size is  0.$''$492.  Log intensity scale; J2000.0 coordinate overlay.
}
\end{figure}


\clearpage


\begin{figure}
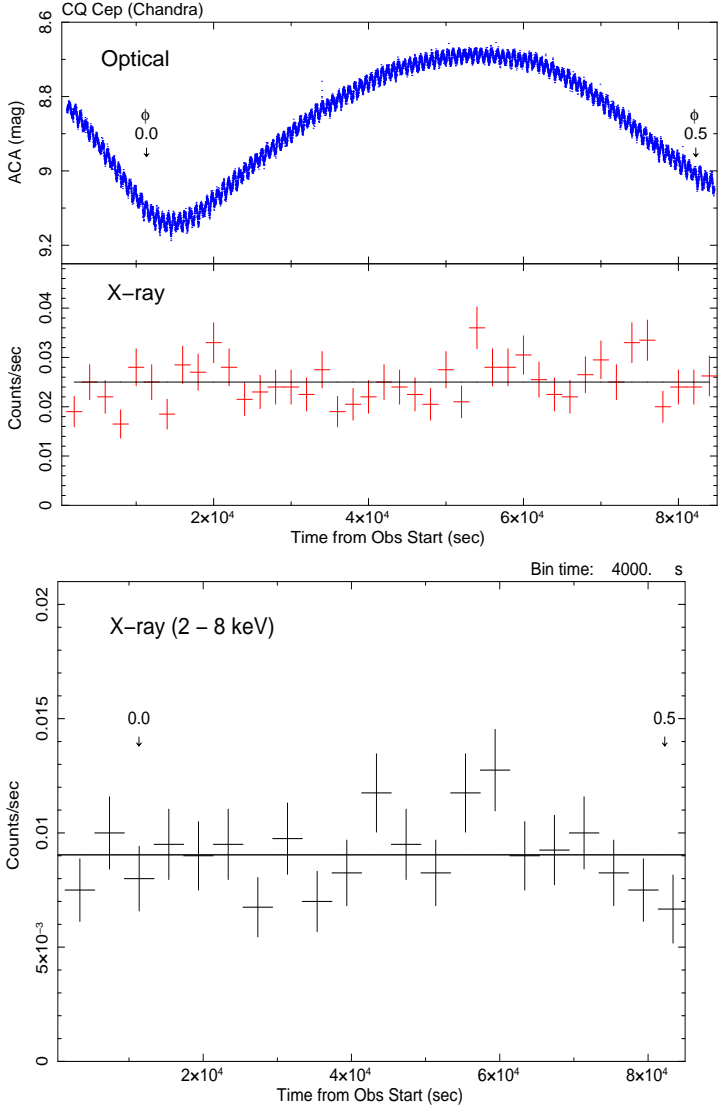

\figurenum{2}
\epsscale{1.0}
\includegraphics*[width=7.4cm,height=10.0cm,angle=-90]{f2t.eps} \\
\includegraphics*[width=7.4cm,height=10.0cm,angle=-90]{f2b.eps} \\
\caption{{\em Top}:~Chandra ACA optical and ACIS-S X-ray light curves of CQ Cep. The optical light
          curve is in units of ACA instrumental magnitude. The small oscillations are
          due to dithering. The optical minimum occurs at $t$ = 14,460 $\pm$ 20 s from
          Obs Start, which converts to MJD = 56370.7063483796 [UTC]. The arrows
          mark the predicted times for O star in front ($\phi$ = 0.0; t = 11,344 s) and WR star
          in front ($\phi$ = 0.5; t = 82,245 s) using the time-of-minimum-light equation  of Demircan et al. (1997).
          The observed minimum occurs $\approx$3.26 ks after the predicted time which can be explained by
          a small difference in the orbital period (Sec. 5.5).
          The X-ray light curve uses
          events  in the 0.2 - 8 keV range and is binned at 2000 s
          intervals. The X-ray error bars are 1$\sigma$ and the solid line shows the
          mean X-ray count rate =
          25.0 ($\pm$ 4.3; 1$\sigma$) c ks$^{-1}$.
~{\em Bottom}:~Chandra hard-band (2 - 8 keV) light curve of CQ Cep binned at
                 4000 s intervals. The X-ray error bars are 1$\sigma$ and the solid line shows the
                 mean X-ray count rate =  9.04 ($\pm$ 1.6; 1$\sigma$) c ks$^{-1}$. Note that
                 the vertical axis scale is different than that of the broad-band X-ray
                 light curve in the top panel. The arrowed
                 phases are the same as in the top panel.
}
\end{figure}


\clearpage


\begin{figure}
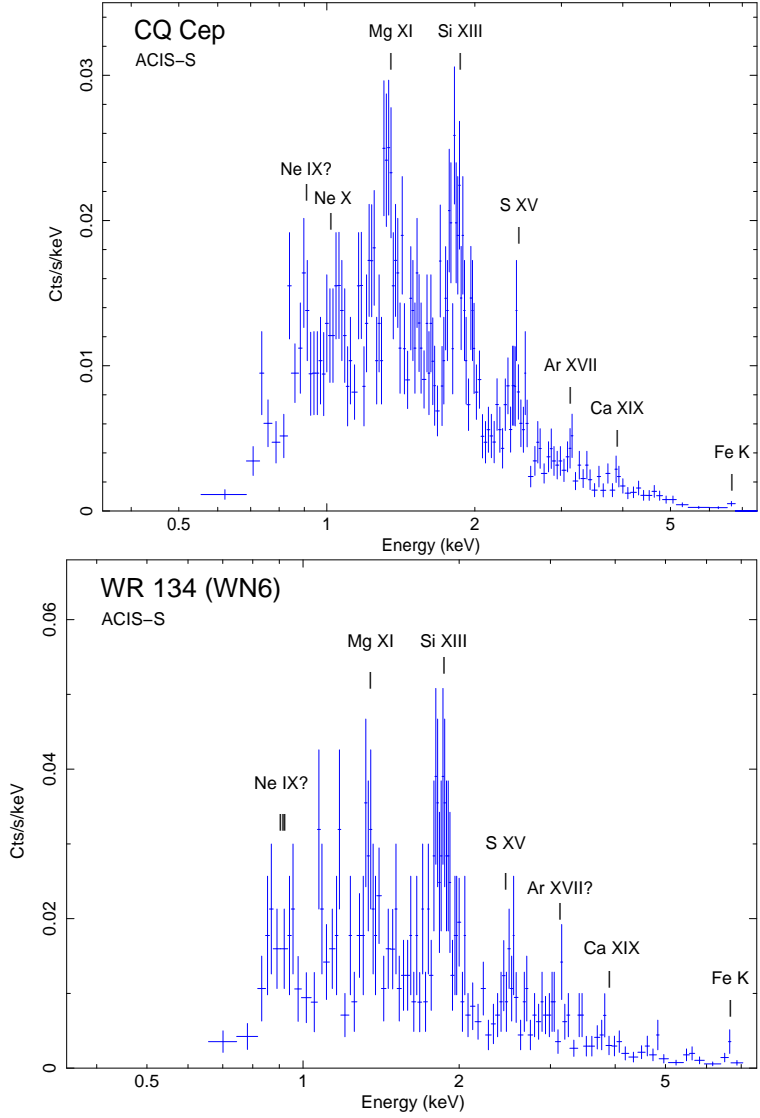

\figurenum{3}
\epsscale{1.0}
\includegraphics*[width=7.38cm,height=10.0cm,angle=-90]{f3t.eps} \\
\includegraphics*[width=7.38cm,height=10.0cm,angle=-90]{f3b.eps}
\caption{~{\em Top}:~Chandra ACIS-S spectrum of CQ Cep (WN6 $+$ OII-Ib) based on
2086 net counts acquired in 79,391 s of usable exposure,
binned to a minimum of  10 counts per bin. The spectrum was extracted from a circular region of
radius 1.$''$8 centered on the source. Possible  emission lines (and blends)  are identified.
~{\em Bottom}:~A comparison Chandra ACIS-S spectrum of the WN6 star WR 134 containing
785 net counts binned to a minimum of 5 counts per bin (Skinner et al. 2010). WR 134
is not known to be a binary. Note that the vertical axis scale is different than that
of CQ Cep in the top panel. At an assumed distance of 1.74 kpc (vdH01) the intrinsic
X-ray luminosity of WR 134 is log L$_{x}$ = 32.66 ergs s$^{-1}$ in the 0.3 - 8 keV
range (Skinner et al. 2010).
}
\end{figure}

\clearpage

\begin{figure}
\figurenum{4}
\epsscale{1.0}
\includegraphics[width=15.0cm,height=5.82cm,angle=0]{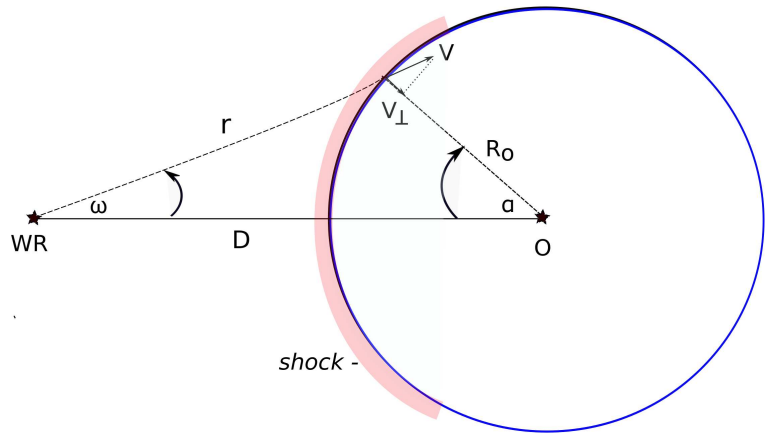}

\caption{Schematic diagram of a close WR$+$O binary showing how the direction of the wind velocity
 component V$_{\perp}$ perpendicular to the shock interface  at the O star surface
 changes as a function of off-axis angle $\omega$. }
\end{figure}

\clearpage


\end{document}